\documentstyle[12pt,epsfig]{article}
                                                                                
\begin{document}                                                                
\date{}
                                                                                
\title{                                                                         
{\vspace{-3cm} \normalsize                                                      
\hfill \parbox{50mm}{DESY 97-058}}\\[25mm]
Leptoquarks and vector-like strong interactions              \\
at the TeV scale                                             \\[12mm]}
\author{ I. Montvay                                          \\
Deutsches Elektronen-Synchrotron DESY,                       \\
Notkestr.\,85, D-22603 Hamburg, Germany}                                       
                                                                                
\newcommand{\be}{\begin{equation}}                                              
\newcommand{\ee}{\end{equation}}                                                
\newcommand{\half}{\frac{1}{2}}                                                 
\newcommand{\rar}{\rightarrow}                                                  
\newcommand{\lar}{\leftarrow}                                                   
                                                                                
\maketitle                                                                      
                                                                                
\begin{abstract} \normalsize
 In a vector-like extension of the minimal standard model with mirror
 fermions leptoquarks can be bound states of fermion-mirror-fermion
 pairs held together by a new strong interaction at the TeV scale.
 The small couplings of leptoquarks to light fermion pairs arise due to
 mixing.
 The large $Q^2$ event excess at HERA and also the high $E_T$ jet excess
 at Tevatron can potentially be explained.
\end{abstract}       

\newpage
\section{Introduction}\label{sec1}
 The recently reported excess of deep-inelastic $e^+p$ events at the
 HERA experiments H1~\cite{H1} and ZEUS~\cite{ZEUS} is an unexpected
 deviation from the predictions of the minimal standard model.
 As has been observed in several theoretical papers~\cite{THEORY},
 a possible explanation can be a resonating leptoquark state in the
 positron-quark channel.
 The existence of such states can be due, among others, to a new strong
 gauge interaction above the presently explored energy range.
 
 In the vector-like extension of the standard model with three mirror
 fermion generations~\cite{FAMILIES} one can naturally implement strong
 interactions at some scale of the order of 1-10 TeV.
 In fact, if the mirror fermions have masses of the order of 200 GeV or
 higher, the renormalization of the corresponding Yukawa couplings
 towards higher energies implies strong interactions at such scales.
 This follows from the renormalization group equations investigated
 in ref.~\cite{GRAND}.
 Since the motivation for the vector-like extension of the standard
 model is based on the difficulty of non-perturbatively defining chiral
 gauge theories, theoretical consistency requires that any new gauge
 interaction should be vector-like.

 In the present letter a simple example based on a strongly interacting
 U(1) Higgs model is exploited.
 For definiteness, let us denote the new charge by $X$.
 The vacuum expectation value breaking this new $\rm U(1)_X$ symmetry
 is assumed to be of the order of several TeV.
 The advantage of U(1) is its simplicity and the fact that it does not
 require the introduction of further new fermionic states.
 Other possibilities based on larger gauge groups are also conceivable,
 but are not considered here.
 The existence of a new strong U(1) interaction in the TeV range has
 been proposed for the explanation of HERA data in the recent paper of
 Babu et al.~\cite{THEORY}.
 Here this idea is implemented in the vector-like extension of the 
 standard model.
 As we shall see, this allows to explain the smallness of the leptoquark
 coupling by relating it to the smallness of the fermion-mirror-fermion
 mixing.

\section{Quantum numbers}\label{sec2}
 In order to introduce the new $\rm U(1)_X$ interactions in the extended
 standard model with three mirror pairs of fermion generations, it is
 advantageous to define a basis of fermion fields where the vector-like
 nature of the gauge interactions becomes explicit.
 Let us denote the fermion fields in an SU(2) doublet by
 $\psi_{L,R}^{(A)}(x)$ and the corresponding mirror fermion fields by
 $\chi_{L,R}^{(A)}(x)$.
 The indices $L,R$ denote chiralities and $A=1,2$ stands for the
 doublet index.
 Since, for instance, $\chi_R^{(A)}(x)$ has the same quantum numbers as
 $\psi_L^{(A)}(x)$, the vector-like gauge couplings act on the
 combinations
\be \label{eq01}                                                                
\rho^{(A)}(x)   \equiv \psi_L^{(A)}(x) + \chi_R^{(A)}(x) \ ,
\hspace{3em}
\sigma^{(A)}(x) \equiv \chi_L^{(A)}(x) + \psi_R^{(A)}(x) \ .
\ee                                                                             
 In the standard model $\rho^{(A)}(x)$ is an SU(2) doublet and
 $\sigma^{(A=1,2)}(x)$ are two SU(2) singlets.
 On the ($\rho,\sigma$)-basis the Yukawa couplings to the standard Higgs
 doublet scalar field are off-diagonal and parity violating.
 The gauge invariant Dirac-masses of these fields are, respectively,
 $\mu_L$ and $\mu_R^{(A=1,2)}$.
 They are responsible for the L-handed, respectively, R-handed
 mixings of fermions with mirror fermions and, therefore, must be
 small in order to be consistent with experiment.
 (For a recent summary of viable mixing schemes and the discussion of
 possible small universality violations due to mixing see 
 ref.~\cite{UNIVERSAL}.)
 The other indices besides the doublet index $A$ are suppressed in
 (\ref{eq01}): in general we have $\rho^{(AcK)}(x)$ and
 $\sigma^{(AcK)}(x)$ with generation index denoted by $K=1,2,3$ and
 the colour index $c$ defined in such a way that $c=1,2,3 \equiv q$
 stand for SU(3) colour and $c=4 \equiv l$ for the corresponding
 lepton.
 On the ($\rho,\sigma$)-basis the
 $\rm SU(3) \otimes SU(2) \otimes U(1)_Y$ gauge interactions are
 explicitly vector-like.
 The new vector-like $U(1)_X$ gauge interaction can generally be
 defined by the charges $X_\rho^{(AcK)}$ and $X_\sigma^{(AcK)}$.
 $\rm SU(3) \otimes SU(2)$ invariance implies that the X-charges are
 the same for $c=1,2,3$ and $X_\rho^{(AcK)}$ is independent of $A$.
 It is plausible that the lightest bound states are X-neutral,
 therefore, in order to produce fermion-antifermion bound states with
 leptoquark quantum numbers, one has to postulate that $X_\rho^{(AcK)}$
 and/or $X_\sigma^{(AcK)}$ be the same for $c=q$ and $l$.
 Therefore, we have the freedom to choose $X_\rho^{(K)}$ and/or
 $X_\sigma^{(AK)}$.

 An important feature of the leptoquark interpretation of the HERA
 data is that leptoquarks must couple, to a very good approximation,
 diagonally to the three generations of quarks and
 leptons~\cite{THEORY}.
 This can be achieved by assuming different X-charges for the three
 fermion generations.
 A possible simple choice is then to assign a single charge value within
 the generations: $X^{(K)} \equiv X_\rho^{(K)}=X_\sigma^{(AK)}$.
 In this case every $\rm U(1)_X$ multiplet can be paired up with any
 other in its generation, giving 256 bound states per generation, if one
 counts isospin and colour components separately.
 Among these states there will be isosinglets, isodoublets and
 isotriplets and as colour multiplets we have singlets, triplets,
 antitriplets and octets.
 It is also possible to introduce models with less states which are 
 still sufficient to explain the large $Q^2$ deviations from the
 standard model observed at HERA:
\begin{itemize}
\item
 {\em $\rho$-model:} taking only $X_\rho^{(K)} \ne 0$ and
 $X_\sigma^{(1K)}=X_\sigma^{(2K)}=0$;
\item
 {\em down-type $\sigma$-model:} taking $X_\sigma^{(2K)} \ne 0$ and
 $X_\sigma^{(1K)}=X_\rho^{(K)}=0$.
\end{itemize}
 Choosing $A=2$ in the second case here is, of course, motivated by the
 required existence of a leptoquark in the $e^+d$-channel, which
 contributes to $e^+p$-scattering.
 In the first model there are isosinglet and isotriplet bound
 states and the number of states per generation is 64.
 In the second one we have only isosinglets and 16 states per
 generation.
 The $\rm SU(3) \otimes SU(2) \otimes U(1)_Y$ quantum numbers of the
 expected light bound states in the down-type $\sigma$-model are
 summarized in table \ref{tab01}.
\begin{table*}[tb]
\begin{center}
\parbox{15cm}{\caption{\label{tab01}\it
 The  $\rm SU(3) \otimes SU(2) \otimes U(1)_Y$ quantum numbers of the
 light fermion-mirror-fermion bound states
 $\Phi_{c_1c_2}^{(K)} \propto
 (\overline{\sigma}^{(2K)}_{c_1} \gamma_5 \sigma^{(2K)}_{c_2})$ in the
 down-type $\sigma$-model.}}
\end{center}
\begin{center}
\begin{tabular}{|c|c||c|c|c|}
\hline
$c_1$  &  $c_2$  &  SU(3)  &  SU(2)  & $Q=\half Y$                  \\
\hline
$e$  &  $e$  & {\bf 1}  & {\bf 1}  &  0                             \\
$d$  &  $e$  & ${\bf \bar{3}}$  & {\bf 1}  &  -2/3                  \\
$e$  &  $d$  & {\bf 3}  & {\bf 1}  &  2/3                           \\
$d$  &  $d$  & ${\bf 1 \oplus 8}$  &  {\bf 1}  &  0                 \\
\hline
\end{tabular} \end{center}
\end{table*}
 In case of the SU(2) quantum numbers one has to have in mind that the
 light fermion states are mixtures of the original chiral fermions with
 their mirror fermion partners.
 This results in an apparent $\rm SU(2)_L$ symmetry violation, if
 compared to the standard model.

 Concerning spin-parity, in analogy with para-positronium, a plausible
 assumption for the lightest fermion-antifermion bound states is
 $J^{PC}=0^{-+}$.
 Other bound states, as $J^{PC}=1^{--}$ vector bosons, are expected to
 have masses in the TeV range.
 The large mass gap between vector and pseudoscalar states might be a
 consequence of an approximate global chiral symmetry, similarly to QCD.
 Nevertheless,
 in general, it is not much known about the dynamical spectrum of
 strongly interacting U(1) Higgs models with fermions.
 Future lattice simulations could help in this respect.

 The qualitative behaviour of the coupling strengths of light
 fermion-antifermion bound states to different types of fermions can
 be inferred from the fermion-mirror-fermion mixing structure.
 In general, one can expect that the strongly bound (pseudo-)scalar
 states are dominantly coupled with a strong Yukawa-coupling of the
 order ${\cal O}(1)$ to their constituent fermions represented by the
 $\rho$-, respectively, $\sigma$-type fields in eq.~(\ref{eq01}).
 The chiral components of fermion ($\psi^{(AcK)}(x)$), respectively,
 mirror fermion ($\chi^{(AcK)}(x)$) fields are mixtures of the mass
 eigenstates, denoted by $\xi^{(AcK)}(x)$ for the light states and
 $\eta^{(AcK)}(x)$ for the heavy states, respectively.
 The mixing relations are:
$$
\psi_{L,R}^{(AcK)}(x) = \xi_{L,R}^{(AcK)}(x) \cos\alpha_{L,R}^{(AcK)}
+ \eta_{L,R}^{(AcK)}(x) \sin\alpha_{L,R}^{(AcK)} \ ,
$$
\be \label{eq02}                                                                
\chi_{L,R}^{(AcK)}(x) =-\xi_{L,R}^{(AcK)}(x) \sin\alpha_{L,R}^{(AcK)}
+ \eta_{L,R}^{(AcK)}(x) \cos\alpha_{L,R}^{(AcK)} \ .
\ee                                                                             
 The mixing angles in the L-handed, respectively, R-handed sector are
 denoted here by $\alpha_L^{(AcK)}$ and $\alpha_R^{(AcK)}$.
 The universality constraints on the $W$- and $Z$-boson couplings imply
 that these mixing angles are small:
 $|\alpha_{L,R}^{(AcK)}| \simeq {\cal O}(10^{-2})$.
 (For the moment we are concentrating on the first generation.
 For the other generations the experimental bounds on the mixing angles
 are weaker.)

 Omitting again colour- and generation-indices as in eq.~(\ref{eq01}),
 for isospin indices $A,B=1,2$ we have in the small mixing angle limit,
 for instance:
$$
\left(\overline{\rho}^{(A)}(x) \gamma_5 \rho^{(B)}(x)\right) \simeq
 \alpha_R^{(A)}\left(\overline{\xi}_R^{(A)}(x)\xi_L^{(B)}(x) \right)
-\alpha_R^{(B)}\left(\overline{\xi}_L^{(A)}(x)\xi_R^{(B)}(x) \right)
$$
\be \label{eq03}
+\alpha_L^{(A)}\left(\overline{\eta}_L^{(A)}(x)\eta_R^{(B)}(x) \right)
-\alpha_L^{(B)}\left(\overline{\eta}_R^{(A)}(x)\eta_L^{(B)}(x) \right)
-\left(\overline{\eta}_R^{(A)}(x)\xi_L^{(B)}(x)\right)
+\left(\overline{\xi}_L^{(A)}(x)\eta_R^{(B)}(x)\right) \ ,
\ee
$$
\left(\overline{\sigma}^{(A)}(x) \gamma_5 \sigma^{(B)}(x)\right) \simeq
 \alpha_L^{(B)}\left(\overline{\xi}_R^{(A)}(x)\xi_L^{(B)}(x) \right)
-\alpha_L^{(A)}\left(\overline{\xi}_L^{(A)}(x)\xi_R^{(B)}(x) \right)
$$
\be \label{eq04}
+\alpha_R^{(B)}\left(\overline{\eta}_L^{(A)}(x)\eta_R^{(B)}(x) \right)
-\alpha_R^{(A)}\left(\overline{\eta}_R^{(A)}(x)\eta_L^{(B)}(x) \right)
+\left(\overline{\eta}_L^{(A)}(x)\xi_R^{(B)}(x)\right)
-\left(\overline{\xi}_R^{(A)}(x)\eta_L^{(B)}(x)\right) \ .
\ee                                                                             
 These relations express the obvious fact that $\rho\rho$- and
 $\sigma\sigma$-bound states are predominantly composed of a light
 fermion and its heavy mirror fermion partner.
 Therefore, the expected Yukawa-couplings in the first generation are
 of the order
\be \label{eq05}                                                                
\lambda_{\Phi\xi\eta} \simeq {\cal O}(1) \ ,
\hspace{3em}
\lambda_{\Phi\xi\xi},\lambda_{\Phi\eta\eta} \simeq {\cal O}(10^{-2}) \ .
\ee                                                                             
 This is in an order of magnitude agreement with the leptoquark coupling
 $\lambda_\Phi \simeq 0.04$ required for the explanation of HERA
 data~\cite{THEORY}.

 In general, a qualitative prediction of models as the ones discussed
 here is, besides leptoquarks, the existence of bound states with
 lepton-antilepton and quark-antiquark quantum numbers (see
 e.g.~table~\ref{tab01}).
 Since at the scale of the new strong interaction the SU(3) colour
 coupling and other standard model couplings are not very strong,
 it is expected that the masses of $\Phi_{ll}$ and $\Phi_{qq}$ are not
 much different from the mass of $\Phi_{lq}$.
 This has interesting implications for the possibility of
 production at high energy colliders.
 At LEP2 the $\Phi_{ll}$ resonance is an interesting candidate, whereas
 at the Tevatron especially the states with colour could be produced.
 Besides the colour triplet $\Phi_{lq}$, also colour octet states of
 the type $\Phi_{qq}$ should be pair produced.
 Their decays produce jet pairs, therefore a $\Phi_{qq}\Phi_{qq}$
 pair decays predominantly to four jets.

\begin{figure}
\begin{center}
\epsfig{file=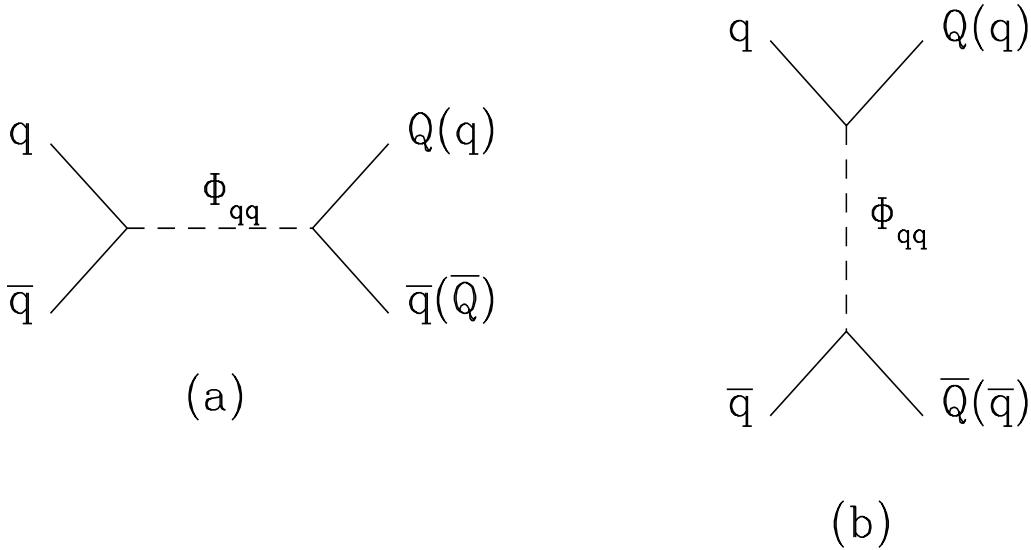,
        width=15.0cm,height=8.0cm,
        bbllx=90pt,bblly=330pt,bburx=590pt,bbury=570pt,
        angle=0}
\end{center}\vspace{0.5em}
\caption{Virtual quark-mirror-quark resonance ($\Phi_{qq}$)
 contributions in $s$-channel (a) and $t$-channel (b) to $q\bar{q}$
 scattering.
 Heavy mirror quarks are denoted by $Q$.}
\label{fig01}
\end{figure}
\section{Consequences for mirror fermions}\label{sec3}
 The existence of leptoquarks and other similar fermion-mirror-fermion
 bound states has important consequences for the phenomenology of mirror
 fermions, too.
 The fermion-mirror-fermion bound state, in general denoted by $\Phi$,
 has lower mass than the sum of masses of the corresponding fermion
 $(f)$ and mirror fermion $(F)$, therefore the $F \to f\Phi$ decay may
 be kinematically possible and goes via the strong coupling
 $\lambda_{\Phi\xi\eta}={\cal O}(1)$.
 If the mass deficit is large enough, this is a fast dominant decay,
 because the decay amplitudes via electroweak vector bosons are
 supressed by mixing angles $\alpha \simeq {\cal O}(10^{-2})$.
 
 As an example, let us consider the decay of the heavy mirror $d$-quark
 ($D$) into a leptoquark ($\Phi_{de}$) plus electron ($e^-$) within the
 down-type $\sigma$-model of table~\ref{tab01}.
 The coupling $\lambda_{\Phi De}$ for $\Phi_{de}(\overline{e}_R D_L)$ is
 expected to be of the order $\lambda_{\Phi De}={\cal O}(1)$.
 The decay width for zero electron mass is
\be \label{eq06}
\Gamma_{D\to\Phi_{de}e} = \frac{M_D\lambda_{\Phi De}^2}{32\pi}
\left(1 - \frac{M_\Phi^2}{M_D^2}\right)^2 \ .
\ee                                                                             
 For $M_D=300\,$GeV and $M_\Phi=200\,$GeV this gives
 $\Gamma_{D\to\Phi_{de}e} \simeq 1\,$GeV.
 This has to be compared to the decay widths to electroweak vector
 bosons which are of the order of MeV \cite{FAMILIES}.
 In addition to these decays, the mirror quarks may also have similar
 decays into quarks plus quark-mirror-quark bound states
 (in general $\Phi_{qq}$, in the model of table~\ref{tab01}
 $\Phi_{dd}$).

 Similarly, $\Phi$ also plays an important r\^ole in the production of
 heavy mirror fermions both in lepton-quark and quark-quark scattering.
 Let us consider, as an example, quark-antiquark scattering relevant
 at Tevatron.
 The important $s$-channel and $t$-channel contributions are shown,
 respectively, by figures~\ref{fig01}a and \ref{fig01}b.
 The amplitudes for $q\bar{q} \to q\bar{Q}$ or $q\bar{q} \to Q\bar{q}$
 are proportional to the couplings
 $\lambda_{\Phi\xi\xi}\lambda_{\Phi\xi\eta} \simeq {\cal O}(10^{-2})$.
 This has to be compared to the production amplitudes via electroweak
 vector bosons which have an additional suppression by an electroweak
 coupling.
 Particularly important is the $t$-channel contribution to the process
 $q\bar{q} \to Q\bar{Q}$ in figure~\ref{fig01}b, because it has only
 couplings $\lambda_{\Phi\xi\eta}={\cal O}(1)$.
 Similar mirror fermion production mechanisms are also effective in
 $e^+e^-$ scattering, where the r\^ole of $\Phi_{qq}$ is played by
 $\Phi_{ee}$ and in the $t$-channel also by $\Phi_{eq}$.
 In $ep$-scattering only $\Phi_{eq}$ is relevant: either in the
 $s$-channel or in the $u$-channel.

 It is an important qualitative feature of this model based on
 fermion-mirror-fermion bound states that the couplings to light
 fermion pairs are suppressed by small mixing angles (see
 eq.~(\ref{eq05}) ).
 This implies that fermion-mirror-fermion bound state exchanges are
 important for heavy mirror fermion production, but are suppressed
 by factors of about ${\cal O}(10^{-3})$-${\cal O}(10^{-4})$ in light
 fermion scattering.

 It is an interesting question, whether the processes mediated by the
 quark-mirror-quark resonances could contribute to the high-$E_T$
 jet cross section excess observed at the Tevatron \cite{CDF}.
 A potentially relevant contribution can arise from the decay products
 of mirror quarks produced by the mechanisms
 with virtual $\Phi_{qq}$ bosons
 shown by figure~\ref{fig01}.
 For instance, the differential cross section of the process
 $q\bar{q} \to Q\bar{Q}$ corresponding to the Feynman graph in
 figure~\ref{fig01}b is
\be \label{eq07}
\frac{d\hat{\sigma}}{d\hat{t}} = 
\frac{\lambda_{\Phi Qq}^4 (M_Q^2-\hat{t})^2}
{64\pi\hat{s}^2 (M_\Phi^2-\hat{t})^2} \ .
\ee                                                                             
 The unknown mirror quark mass $M_Q$ is expected to be above 200 GeV, if
 the present interpretation of the observed high $Q^2$ anomaly at HERA
 by leptoquark resonances is correct.
 As discussed above, the coupling $\lambda_{\Phi Qq}$ is expected to
 be of the order $\lambda_{\Phi Qq}={\cal O}(1)$.
 Compared to the heavy quark production parton cross sections in QCD
 \cite{QCDQQ} the ratio of total parton cross sections is roughly given
 by the ratio of couplings $(\lambda_{\Phi Qq}/g_s)^4$, where $g_s$ is
 the QCD coupling.
 For instance, for $\lambda_{\Phi Qq}/g_s \simeq 3$ we have a total
 parton cross section resulting from (\ref{eq07}) which is by two orders
 of magnitude larger than those from heavy quark pair production
 processes in QCD.
 This is large enough to give a significant contribution to the single
 jet inclusive cross section above 200 GeV and can compete with
 interpretations based on modifications of the parton distributions
 (see, for instance, \cite{PARTON}).
 However, the cascade decays $Q \to \Phi_{qq}+jet \to jet+jet+jet$
 produce multi-jet signature characteristics, which distinguish these
 final states from usual QCD ones.
 In QCD-like final states some virtual loop contributions involving
 $Q$, $\Phi_{qq}$ and the coupling $\lambda_{\Phi Qq}$ could also be
 relevant.
 Another contribution to high-$E_T$ can come from a process as in
 figure \ref{fig01}a with on-shell $\Phi_{qq}$ resonance and virtual
 $Q$.
 This would contribute to the cross-sections of W or Z plus jets.
 The evaluation of these different possibilities requires a detailed
 study, which goes beyond the scope of this letter.

 In summary, the assumption of a new strong interaction at the TeV scale
 in the vector-like extension of the standard model with mirror fermions
 can produce resonances in fermion-mirror-fermion channels.
 These new states can potentially explain the recently observed high
 $Q^2$ and high $E_T$ event excesses compared to the predictions of the
 minimal standard model.
 The small mixing between fermions and mirror fermions naturally
 leads to small couplings of these new resonant states to light fermion
 pairs.

\vspace{10mm}                                                    
{\large\bf Acknowledgements} 

\vspace{5mm}\noindent
 It is a pleasure to thank W. Buchm\"uller and P. Zerwas for
 discussions.


\end{document}